\documentclass[aps,prl,twocolumn,superscriptaddress]{revtex4-1}
\usepackage{graphicx}
\usepackage{amsmath}
\usepackage{amssymb}
\usepackage{graphicx}\usepackage{afterpage}
\usepackage{amsfonts}
\usepackage{amssymb}
\usepackage{graphicx}
\usepackage{braket}
\usepackage{textcomp}

\newcommand{\Ea}{\ensuremath{{\cal E}_1}}
\newcommand{\Eb}{\ensuremath{{\cal E}_2}}
\newcommand{\Ec}{\ensuremath{{\cal E}_3}}

\newcommand{\Er}{\ensuremath{{\cal E}_{\rm R}}}
\newcommand{\Eo}{\ensuremath{{\cal E}_{1,2,3}}}

\begin{document}

\title{Impact of phonons on dephasing of  individual excitons\\ in deterministic quantum dot microlenses}

\author{T.~Jakubczyk}
\email{tomasz.jakubczyk@neel.cnrs.fr} \affiliation{Univ. Grenoble
Alpes, F-38000 Grenoble, France} \affiliation{CNRS, Institut
N\'{e}el, "Nanophysique et semiconducteurs" group, F-38000 Grenoble,
France}

\author{V.~Delmonte}
\affiliation{Univ. Grenoble Alpes, F-38000 Grenoble, France}
\affiliation{CNRS, Institut N\'{e}el, "Nanophysique et
semiconducteurs" group, F-38000 Grenoble, France}

\author{S.~Fischbach}
\affiliation{Institut f\"{u}r Festk\"{o}rperphysik, Technische
Universit\"{a}t Berlin, Hardenbergstra{\ss}e 36, D-10623 Berlin,
Germany}

\author{D.~Wigger}
\email[]{d.wigger@wwu.de} \affiliation{Institut f\"{u}r
Festk\"{o}rpertheorie, Universit\"{a}t M\"{u}nster, 48149
M\"{u}nster, Germany}

\author{D.~E.~Reiter}
\affiliation{Institut f\"{u}r Festk\"{o}rpertheorie, Universit\"{a}t
M\"{u}nster, 48149 M\"{u}nster, Germany}

\author{Q.~Mermillod}
\affiliation{Univ. Grenoble Alpes, F-38000 Grenoble, France}
\affiliation{CNRS, Institut N\'{e}el, "Nanophysique et
semiconducteurs" group, F-38000 Grenoble, France}

\author{P.~Schnauber}
\affiliation{Institut f\"{u}r Festk\"{o}rperphysik, Technische
Universit\"{a}t Berlin, Hardenbergstra{\ss}e 36, D-10623 Berlin,
Germany}

\author{A.~Kaganskiy}
\affiliation{Institut f\"{u}r Festk\"{o}rperphysik, Technische
Universit\"{a}t Berlin, Hardenbergstra{\ss}e 36, D-10623 Berlin,
Germany}

\author{J.-H.~Schulze}
\affiliation{Institut f\"{u}r Festk\"{o}rperphysik, Technische
Universit\"{a}t Berlin, Hardenbergstra{\ss}e 36, D-10623 Berlin,
Germany}

\author{A.~Strittmatter}
\affiliation{Institut f\"{u}r Festk\"{o}rperphysik, Technische
Universit\"{a}t Berlin, Hardenbergstra{\ss}e 36, D-10623 Berlin,
Germany}

\author{S.~Rodt}
\affiliation{Institut f\"{u}r Festk\"{o}rperphysik, Technische
Universit\"{a}t Berlin, Hardenbergstra{\ss}e 36, D-10623 Berlin,
Germany}

\author{W.~Langbein}
\affiliation{Cardiff University School of Physics and Astronomy, The
Parade, Cardiff CF24 3AA, United Kingdom}

\author{T.~Kuhn}
\affiliation{Institut f\"{u}r Festk\"{o}rpertheorie, Universit\"{a}t
M\"{u}nster, 48149 M\"{u}nster, Germany}

\author{S.~Reitzenstein}
\email[]{stephan.reitzenstein@physik.tu-berlin.de}
\affiliation{Institut f\"{u}r Festk\"{o}rperphysik, Technische
Universit\"{a}t Berlin, Hardenbergstra{\ss}e 36, D-10623 Berlin,
Germany}

\author{J.~Kasprzak}
\email[]{jacek.kasprzak@neel.cnrs.fr} \affiliation{Univ. Grenoble
Alpes, F-38000 Grenoble, France} \affiliation{CNRS, Institut
N\'{e}el, "Nanophysique et semiconducteurs" group, F-38000 Grenoble,
France}

\begin{abstract}
Optimized light-matter coupling in semiconductor nanostructures is a
key to understand their optical properties and can be enabled by
advanced fabrication techniques. Using in-situ electron beam
lithography combined with a low-temperature cathodoluminescence
imaging, we deterministically fabricate microlenses above selected
InAs quantum dots (QDs) achieving their efficient coupling to the
external light field. This enables to perform four-wave mixing
micro-spectroscopy of single QD excitons, revealing the exciton
population and coherence dynamics. We infer the temperature
dependence of the dephasing in order to address the impact of
phonons on the decoherence of confined excitons. The loss of the
coherence over the first picoseconds is associated with the emission
of a phonon wave packet, also governing the phonon background in
photoluminescence (PL) spectra. Using theory based on the
independent boson model, we consistently explain the initial
coherence decay, the zero-phonon line fraction, and the lineshape of
the phonon-assisted PL using realistic quantum dot geometries.
\end{abstract}

\maketitle

Owing to the progress in the semiconductor growth, the
self-assembled quantum dots (QDs) offer nowadays optimal quality of
the residing exciton transitions\,\cite{BorriPRL01} with enhanced
emission efficiency\,\cite{CurtoSci10,ClaudonNatPhot10, MaOL15} and
close to ideal quantum optical properties\,\cite{DingPRL16}.
Forthcoming applications emerging from combining QDs and
nanophotonics -- such as, quantum light sources in on-chip photonic
networks -- call for scalability and deterministic QD positioning.
In this regard, in-situ electron beam
lithography\,\cite{donatiniPatent10,DonatiniNanotech10} (EBL) has
evolved into a suited technique for the deterministic fabrication of
quantum light sources\,\cite{GschreyNatComm15, KaganskiyRSI15}. When
combined with a low temperature cathodoluminescence imaging, EBL
permits to sculpture microlenses above individual QDs, enhancing
collection efficiency over a broad spectral range.

Here, using four-wave mixing (FWM) micro-spectroscopy we reveal
coherences of \emph{single} QDs, deterministically embedded in
microlenses realized by EBL. The resulting optical signals in these
nanophotonic structures exhibit an enhanced signal to noise ratio.
This enables us to infer the impact of acoustic phonons on the
coherence dynamics of individual QDs. In particular, we report on
the exciton zero phonon line (ZPL) dephasing close to the radiative
limit in a single QD at 5\,K, distinguishing it from its spectral
wandering. Phonons are known to play a crucial role in the optical
control of QDs \cite{ReiterJPC14,RamsaySST10}. With increasing
temperature, we observe an increasing impact of phonon-induced
dephasing \cite{BorriPRB05} owing to the polaron formation and wave
packet emission \cite{KrugelPRB07,WiggerJPC14} and a broadening of
the homogenous width $\gamma$, attributed to a quadratic coupling
between carriers and acoustic phonons
\cite{MuljarovPRL04,MachnikowskiPRL06}. Single QD micro-spectroscopy
permits to associate the measured dephasing during the polaron
formation with the spectral shape of phonon-assisted transitions,
here accessed via photoluminescence (PL). We thus go beyond the FWM
experiments performed on QD ensembles\,\cite{BorriPRL01,
BorriPRB05}, and we consistently explain the initial FWM decay, the
zero-phonon line (ZPL) fraction, and the lineshape of the
phonon-assisted PL using a realistic QD geometry. Additionally, the
optical parameters of QDs - i.e. dephasing, lifetime, dipole
moments, phonon coupling - are to some extent averaged in QD
ensemble measurement due to stochastic distribution of their shapes
and alloy composition. This issue is naturally overcome in a single
QD spectroscopy carried out here.

When performing FWM on single emitters on a simple planar structure
one is confronted with a huge ratio between the resonant background
(typically $10^6-10^8$ in the field, and $10^5$ when assisted with
high quality anti-reflection coatings \cite{LangbeinOL06}) and the
induced FWM. We have recently shown that, using suitable photonic
nanostructures \cite{MermillodPRL16,FrasNatPhot16}, one can boost
the experimental sensitivity by bringing a large amount of the field
amplitude to the vicinity of a QD. This enhances its interaction
with the excitonic dipole, and hence reduces the required external
power constituting the background. Furthermore, using nanophotonic
devices, the FWM is efficiently collected by the detection optics,
avoiding the total internal reflection affecting planar structures:
assuming the gain in the collection efficiency $\eta$ and an
$n$-time enhancement of the local excitation intensity, the FWM
amplitude is increased by $\sqrt{\eta n^3}$, while maintaining the
external power of exciting laser pulses $\Eo$.

In contrast to our previous works, the EBL overcomes the issue of a
low yield of optimally functioning devices, when patterning the
sample containing randomly distributed QDs. Microlenses processed
with EBL can be defined deterministically, spatially matched to QDs
with about 30\,nm alignment accuracy \cite{GschreyJVST2015},
combined with frequency matching guaranteed by their broadband
operation beyond 100\,nm.

\begin{figure}[t]
\includegraphics[width=1.02\columnwidth]{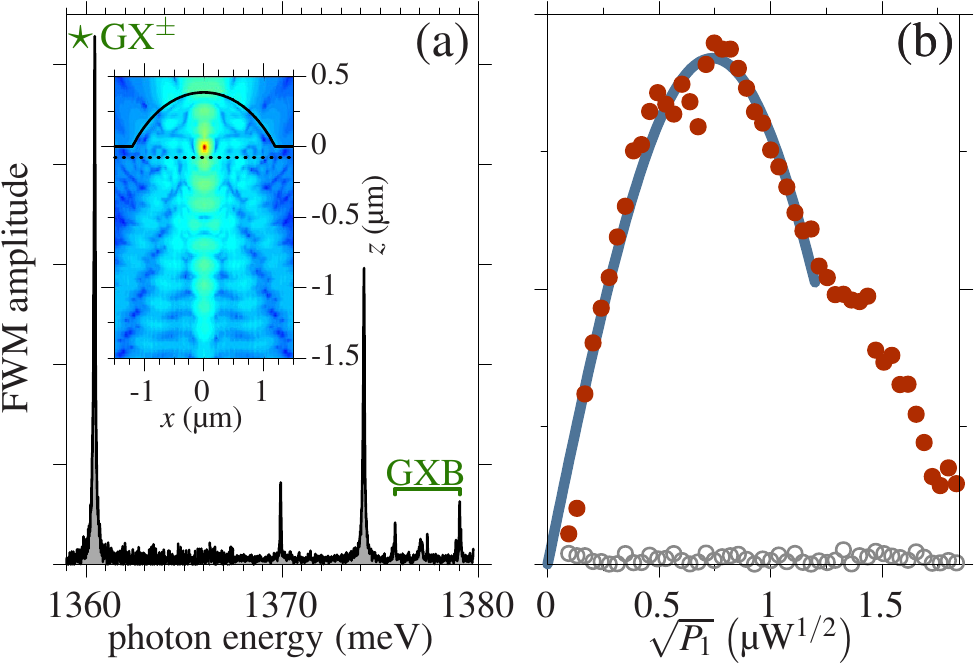}
\caption{(a) Spectrally-resolved FWM amplitude generated by a few QD
excitons embedded in a lens structure. The targeted QD trion
(GX$^\pm$) is labeled with the green $\star$. The horizontal bar
indicates a neutral exciton-biexciton system (GXB) in a QD located
at the lens periphery. Inset: Calculated distribution of the
near-field intensity for the QD-lens structure. The
semiconductor-air interface is shown by the solid black line and the
DBR starts below the dashed line. (b) Spectrally-integrated FWM
amplitude of a trion in the target QD as a function of the pulse
area $\theta_1\propto\sqrt{P_1}$ of $\Ea$. The blue line shows the
fit to the expected $|\sin(\theta_1/2)|$ amplitude dependence of the
FWM. \label{fig:Fig1}}
\end{figure}

Microlenses with a height of 0.35\,\textmu m and 2\,\textmu m
diameter have been etched, as described in detail in
Ref.~\cite{KaganskiyRSI15}, so as to create the hot spot of the
excitation and detection mode field exactly at the QD plane, located
between the underneath Bragg reflector and the microlens surface
\cite{GschreyNatComm15} - see inset in Figure\,\ref{fig:Fig1}a. We
use InAs QDs grown by metalorganic chemical vapor deposition
\cite{ProhlAPL13}. As a result of the high light-extraction
efficiency \cite{SchlehahnAPL16} of around 30\,\%\ in our devices,
we routinely note a bright QD photoluminescence, with
spectrally-integrated count rate of 200\,kHz below the PL saturation
(not shown). For comparison, up to 1\,kHz PL count rates are
typically observed in our setup for high quality QDs embedded in
planar samples.

For the FWM spectroscopy we use radio-frequency acousto-optic
deflectors providing frequency shifts of $\Omega_{1,2,3}$ for $\Eo$,
respectively. FWM signals, in the lowest order proportional to
$\Ea^{\ast}\Eb\Ec$ ($\ast$ denoting complex conjugate), are then
detected by performing optical heterodyning. To select the required
heterodyne beat component, we interfere the reflected field with a
frequency shifted reference field $\Er$. Spectrally-resolved
interferograms are recorded by a CCD camera
\cite{LangbeinOL06,FrasNatPhot16} and analyzed by spectral
interferometry to retrieve amplitude and phase of the signal. In
Figure\,\ref{fig:Fig1}a we show an exemplary two-pulse FWM spectrum,
(driving with $\Ea$ and $\Eb$ and detecting at the heterodyne
frequency $2\Omega_2-\Omega_1$), over a range of 25\,meV displaying
several excitonic transitions. Owing to a strongly improved
excitation and collection compared to planar structures
\cite{KasprzakNJP13}, we here achieve a gain in the signal-to-noise
ratio of the measured FWM amplitude by two-orders of magnitude.
Exciton-biexciton pairs can be identified (an example is denoted
with a green bar above 1375\,meV) by employing FWM polarization and
delay selection rules \cite{MermillodOptica16} (not shown). A
similar FWM signal was found on another investigated microlens,
pointing toward the deterministic character and high-quality of the
EBL nano-processing platform.

In the following, we study the dominating transition at 1360.4\,meV
labeled as {\large$\star$} in Figure\,\ref{fig:Fig1}a. This is to
minimize the time required to perform following FWM sequences and
thus to avoid drifts. We observe no FWM at $\tau_{12}<0$, showing
that it stems from a charged exciton (trion) transition GX$^{\pm}$
\cite{MermillodOptica16}. To illustrate the enhanced in-coupling of
$\Eo$ offered by the microlenses, in Figure\,\ref{fig:Fig1}b we
present the FWM amplitude as a function of the pulse area
$\theta_1\propto\sqrt{P_1}$, where $P_1$ represents the intensity of
$\Ea$, while $P_2$ is fixed to $1.5$\,\textmu W. The FWM signal
displays a Rabi rotation following the expected $|\sin(\theta_1/2)|$
dependence \cite{PattonPRL05} with the first maximum at
$\sqrt{P_1}=0.75$\,\textmu W$^{1/2}$, corresponding to a pulse area
of $\theta_1=\pi/2$. For higher intensities the FWM signal deviates
from this behavior.

\begin{figure}[t]
\includegraphics[width=1.02\columnwidth]{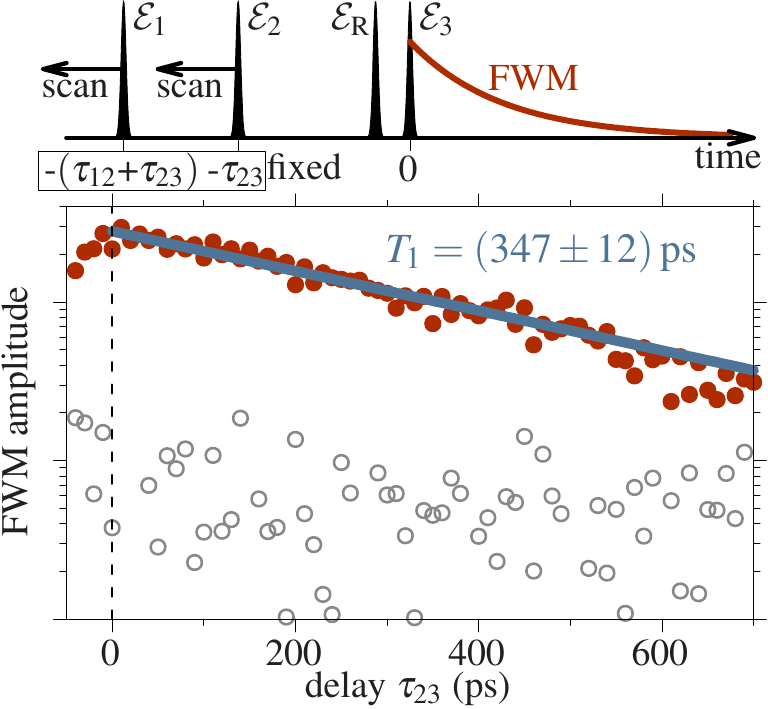}
\caption{Top: Three-pulse sequence employed to measure the trion
population dynamics. $\Ea$ and $\Eb$, having a  delay of
$\tau_{12}=20\,$ps, create the trion population and are jointly
advanced in time, such that the FWM triggered by $\Ec$ probes the
population decay via the $\tau_{23}$ dependence. Bottom: Measurement
yielding the exciton lifetime $T_1=(347\pm 12)\,$ps. The noise level
is indicated by open circles.\label{fig:Fig2}}
\end{figure}

To measure the exciton density lifetime $T_1$, we employed the
three-pulse FWM, where the signal is detected at
$\Omega_3+\Omega_2-\Omega_1$, as a function of the second delay
$\tau_{23}$, displayed in Figure\,\ref{fig:Fig2}. From its
exponential decay we determine the lifetime
\cite{MermillodPRL16,FrasNatPhot16} to $T_1=(347\pm 12)\,$ps at
$T=5$\,K. Such a rather short lifetime, compared to about 1\,ns
typically observed \cite{ThomaPRL16} in these structures, is
attributed to the selectivity of the FWM technique favoring
particularly bright QDs with high dipole moment and thus displaying
fast population decay dynamics (less intense transitions in
Figure\,\ref{fig:Fig1}a are expected to exhibit longer $T_1$).
Additionally, the radiative lifetime is slightly shortened due to a
Purcell effect of the microlenses \cite{KaganskiyRSI15}.

\begin{figure}[t]
\includegraphics[width=1.02\columnwidth]{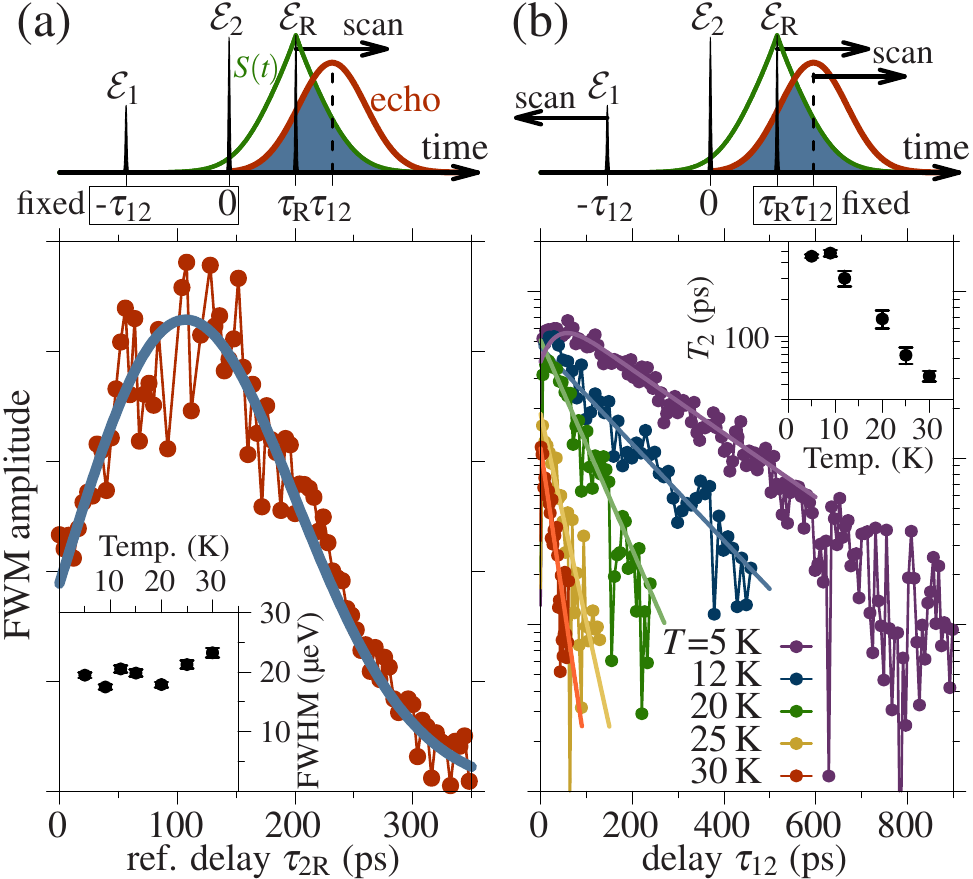}
\caption{ (a) Top: Two-pulse sequence applied to probe the echo
profile. Bottom: Integrated FWM amplitude versus $\tau_{2{\rm R}}$
at 5\,K revealing the Gaussian echo with a temporal width yielding
$\sigma$; the theoretical fit is given by the solid line. Inset:
Inhomogeneous broadening $\sqrt{8{\rm ln}(2)}\sigma$ retrieved from
the echo temporal width for different temperatures. (b) Top:
Two-pulse sequence applied to probe the trion dephasing. Bottom:
Measured FWM amplitude as function of the delay $\tau_{12}$,
yielding coherence dynamics for different temperatures; theoretical
fits as solid lines. Inset: Dephasing time $T_2$ as a function of
temperature. \label{fig:Fig3}}
\end{figure}

We now turn to the measurement of coherence dynamics as a function
of temperature, to determine the impact of the phonon-interaction on
the exciton dephasing. The FWM transient, generated in our
time-averaged and multi-repetition heterodyne experiment, is emitted
after the arrival of $\Eb$. It is expected to exhibit a Gaussian
echo \cite{KasprzakNJP13, MermillodPRL16}, owing to a Gaussian
spectral wandering of standard deviation $\sigma$, with the maximum
at $t=\tau_{12}$ and temporal full width at half maximum (FWHM) of
$\hbar\sqrt{8\ln(2)}/\sigma$. To retrieve $\sigma$, we apply the
pulse sequence depicted in Figure\,\ref{fig:Fig3}a, namely we keep
$\tau_{12}=110\,$ps fixed, while scanning the delay $\tau_{2\rm{R}}$
between $\Eb$ and $\Er$. As such, the temporal sensitivity of the
experiment $S(t)$ (green curve centered around $\Er$), originating
from the finite spectral resolution of the spectrometer, is scanned
through a broad FWM transient. A measurement of $S(t)$ is given in
the Supplementary Figure\,S1. The FWM integrated overlap between
$S(t)$ and the echo plotted against $\tau_{2\rm{R}}$ in
Figure\,\ref{fig:Fig3}a indeed reproduces a Gaussian form, with the
expected maximum at $\tau_{12}=110$\,ps and $\sigma=8.2$\,\textmu
eV. The measured inhomogeneous broadening (FWHM)
$\sqrt{8\ln(2)}\sigma$ is plotted versus temperature in the inset,
where we find that it varies only marginally within the investigated
temperature range.

To extract the ZPL dephasing rate $\gamma=1/T_2$ we measure the
time-integrated FWM amplitude as a function of $\tau_{12}$. For a
fixed $\tau_{2\rm{R}}$, the echo moves through $S(t)$ when varying
$\tau_{12}$. This has previously been compensated by correcting the
signal in the time domain by $S(t)$
\cite{KasprzakNJP13,MermillodPRL16,FrasNatPhot16}. However, for
sufficiently large $\tau_{12}$, the echo is generated at times not
accessible via $S(t)$, such that the signal cannot be retrieved via
spectral interference. Here, this issue is overcome by
simultaneously increasing $\tau_{2\rm{R}}$ towards positive times
when increasing $\tau_{12}$, such that $S(t)$ probes the same
time-portion of the echo for every $\tau_{12}$, as depicted in
Figure\,\ref{fig:Fig3}b. We initially set $\tau_{2\rm{R}}=-70\,$ps
to assure the time ordering between $\Er$ and FWM, as required to
perform spectral interferometry. The resulting FWM amplitude for
several temperatures is presented in Figure\,\ref{fig:Fig3}b.

After the echo has developed for $\tau_{12}>150\,$ps, the decay of
the signal is given by a single exponential, yielding the dephasing
time $T_2$. At low temperature the latter reaches
$T_2\approx1.3T_1$, close to the radiative limit ($T_2=2T_1$), in
spite of the significant inhomogeneous broadening. As shown in the
inset, with increasing temperature, $T_2$ shortens rapidly
consistently with previous measurements on ensembles
\cite{BorriPRB05} and more recent complementary approaches employing
photon-correlation techniques \cite{ThomaPRL16}. The dominant term
in the electron-phonon coupling in semiconductors is linear in the
lattice displacement, i.e., it is linear in the phonon creation and
annihilation operators. For the present case of a QD excited at the
lowest exciton transition, which represents a two-level system, this
reduces to the independent boson model \cite{Mahan2000}. This model,
for the 3D acoustic phonon density of states, provides a band of
phonon assisted transitions and an unbroadened ZPL. The finite width
of the ZPL and its temperature dependence are explained by phonon
processes, which are of second order in the phonon operators, and
may originate from virtual transitions to higher excitonic states
\cite{MuljarovPRL04} or phonon anharmonicities
\cite{MachnikowskiPRL06}, which are not included in our model.

\begin{figure}[t]
\includegraphics[width=1.02\columnwidth]{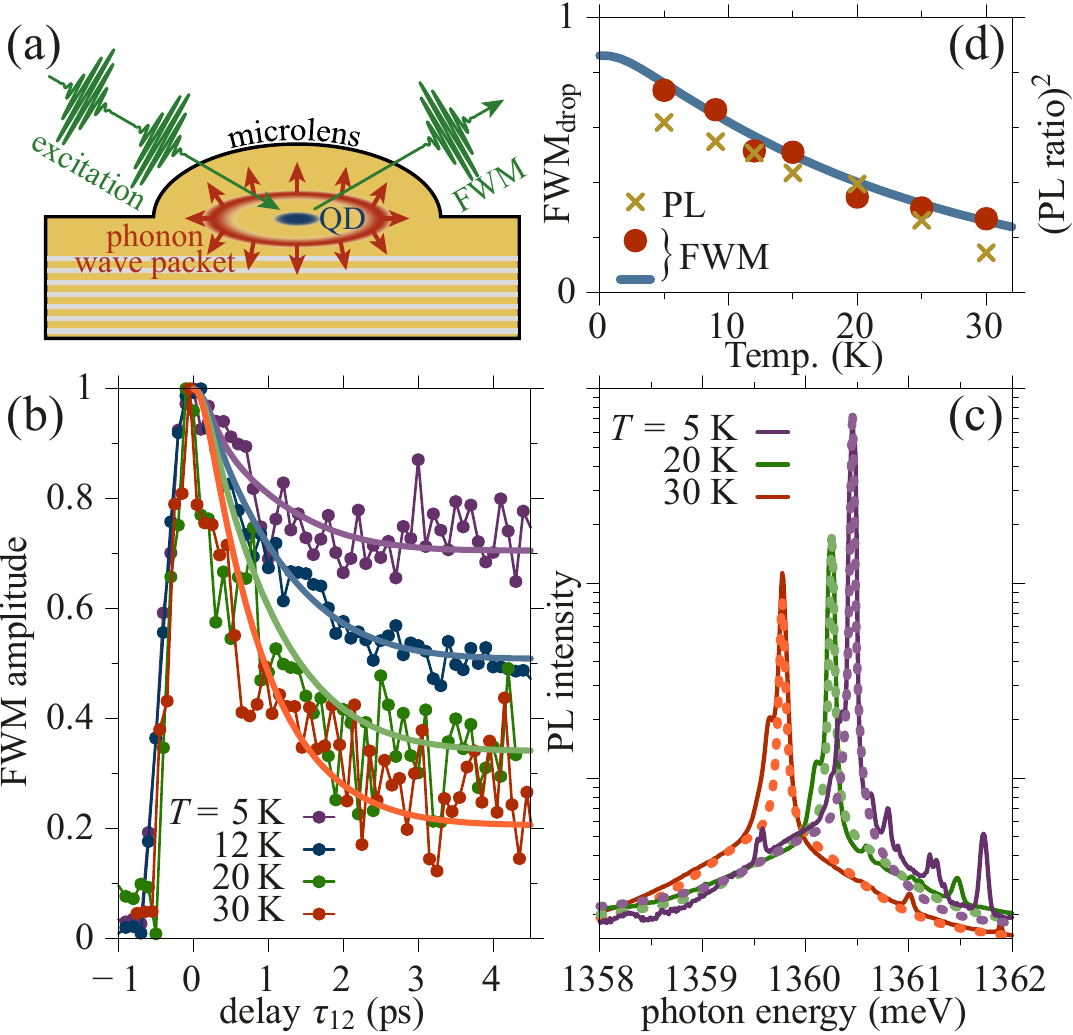}
\caption{ (a)\,Cartoon depicting propagation of a phonon packet from
a QD after its excitation with a short, femto-second pulse.
(b)\,Two-pulse time-integrated FWM amplitudes for initial delays
$\tau_{12}$. The FWM amplitudes at different temperatures (see
legend) are normalized at $\tau_{12}=0$ to show the phonon-induced
dephasing. (c)\,PL spectra for different temperatures. Solid lines:
experimental data; dashed lines: theoretical curves. (d)\,Final FWM
values after the initial decay (red cirles) as function of
temperature along with the theoretical calculation (blue line) (cf.
panel b). Additional temperatures for (b) and (c) are shown in the
Supplementary Figure\,S2. $Z^2$ estimated from temperature dependent
PL spectra are given by gold crosses. \label{fig:Fig4}}
\end{figure}

Figure\,\ref{fig:Fig3}b also reveals a pronounced decrease of the
initial FWM amplitude at $\tau_{12}=0$ with increasing temperature,
such that the signal cannot be measured beyond $T=30$\,K. This rapid
initial decrease is attributed to the phonon-induced dephasing
caused by the linear coupling to phonons, which dominates the
short-time behavior of the signal. To analyze this effect we measure
the coherence dynamics on a picosecond time scale. The results are
shown in Figure\,\ref{fig:Fig4}. The linear coupling describes the
fact that the equilibrium positions of the lattice ions in the
presence of an exciton are different from their values in the
absence of an exciton, i.e., a polaron is formed. When an exciton is
abruptly generated by a femto-second pulse, the quick formation of
the polaron is accompanied by the emission of a phonon wave packet
\cite{WiggerJPC14, krummheuerPRB05}, traveling through the QD volume
and then through the surrounding lattice with the sound velocity of
about $5\,$nm/ps, as illustrated in Figure\,\ref{fig:Fig4}a. Once
the wavepacket has left the QD, the phonon-assisted transitions have
dephased and are not further contributing to the FWM. In the
$\tau_{12}$-dependence of the FWM signal this manifests itself as a
fast decay on a timescale of about $2$\,ps \cite{KrummheuerPRB02}
clearly revealed in Figure\,\ref{fig:Fig4}b.

The final value after the initial drop FWM$_{\rm drop}$ is plotted
as a function of temperature in Figure\,\ref{fig:Fig4}d. With
increasing temperature the phonon coupling becomes more effective
and, accordingly, already at $T=30$\,K the coherence decays by a
factor of 5 during the first $5$\,ps. This explains why the initial
value of the signal seen in Figure\,\ref{fig:Fig3}b, where this
initial decay is not resolved, rapidly decays with increasing
temperatures.

In the spectral domain, the initial decay is associated with a broad
phonon background around a ZPL \cite{BesombesPRB01,KrummheuerPRB02}.
This is seen in the PL spectra taken from the same QD, shown in
Figure\,\ref{fig:Fig4}c. For low temperatures the background is
asymmetric reflecting the dominance of phonon emission processes
over absorption processes, while at higher temperatures, when the
thermal occupation of the involved phonons becomes much larger than
one, the phonon background becomes symmetric.

For the theoretical modeling of the signals we employ the standard
model of a QD coupled to acoustic phonons via the pure dephasing
mechanism
\cite{BesombesPRB01,KrummheuerPRB02,ForstnerPRL03,RamsaySST10,ReiterJPC14},
which has been proven to successfully describe a variety of optical
phenomena in single QDs and QD ensembles. For this model, exact
analytical formulas for linear and nonlinear optical signals after
excitation with an arbitrary series of short laser pulses can be
obtained within a generating function formalism
\cite{VagovPRB02,AxtPRB05}. To be specific, we model the QD trion
transition as a two-level system, which is coupled via deformation
potential coupling to longitudinal acoustic (LA) phonons with a
linear dispersion relation. Assuming an approximately harmonic
confinement potential, we take Gaussian-shaped wave functions for
electrons and holes with electron localization lengths $a_r$ in the
in-plane direction and $a_z$ in the out-of-plane direction, i.e., we
model a lens-shaped QD and treat the exciton wave function as a
product of electron and hole wave functions.

The results of our calculations are shown as solid lines in
Figure\,\ref{fig:Fig4}b and as dashed lines in
Figure\,\ref{fig:Fig4}c. We use electron localization lengths
$a_r=8$\,nm and $a_z=1$\,nm (the respective hole localization
lengths are scaled by 0.87) and assume GaAs parameters. We find an
excellent agreement between theory and experiment for both the FWM
signals and the PL spectra over the whole range of temperatures,
however for an increased phonon coupling strength. Specifically, to
achieve agreement with both the FWM signal and phonon background in
PL simultaneously, we had to increase the phonon coupling constant
by a factor of 1.5. This we model by increased deformation
potentials $D^{\rm e}=10.5$\,eV and $D^{\rm h}=-5.25\,$eV, with
respect to the standard parameters $D^{\rm e} = 7.0$\,eV and $D^{\rm
h}=-3.5$\,eV, which have previously been used to quantitatively
describe various optical signals from QD structures
\cite{ReiterJPC14,VagovPRB04,QuilterPRL15}. A more detailed
discussion of the role of the parameters for the FWM signals and PL
spectra can be found in the Supplemental Information. While we do
not have a definite explanation, we note that similarly higher
values for the deformation potentials can also be found in the
literature \cite{Adachi92}, and an increased deformation potential
has also been used to explain mobilities in a 2DEG
\cite{gorczycaPRB92,kawamura1990temperature,scholzJAP95}. One
explanation could be that in the present sample there are additional
mechanisms like piezo-electric coupling, which are usually
negligible but contribute here, e.g., because of an increased
spatial separation of electron and hole wave functions
\cite{krummheuerPRB05}, leading to an effective increase of the
coupling described by larger deformation potentials.

Finally, it is instructive to compare the FWM initial decay with the
ZPL weight in PL ($Z$), defined as fraction of ZPL in the total
absorption spectrum \cite{BorriPRB05}. Within the formalism
describing our experiment we obtain FWM$_{\rm drop}\propto Z^2$. We
approximate $Z$ as the PL ratio between the measured ZPL and the
entire PL, including the phonon background. In spite of finite
spectral resolution and significant $\sigma$, for all considered
temperatures we obtain close agreement between FWM$_{\rm drop}$ and
$Z^2$ independently estimated from PL, as shown in
Figure\,\ref{fig:Fig4}d.

In this Article, we have employed EBL to deterministically embed QDs
within microlenses, providing a convenient nanophotonic platform to
perform coherent nonlinear spectroscopy of individual QDs. Microlens
structures enabled to efficiently penetrate across the dielectric
boundary and to tightly focus the light field at the QD location,
which has been exploited to perform FWM micro-spectroscopy. We have
measured and modeled the role of acoustic phonons on the coherence
of single QD excitons, in particular corroborating signatures of
single phonon wave packet emission in FWM and PL. Our fundamental
studies, aiming to understand the complex interplay between charges
and lattice vibrations, are at the heart of condensed matter optics.
They are relevant for a large class of individual emitters in
solids, like epitaxial and colloidal QDs or colour centres in
diamond, or emerging QD-like emitters in transition-metal
dichalcogenides\,\cite{KernAdvMat16, KumarOptica16, HeOptEx16}. Our
findings are also pertinent for ultrafast nonlinear nanophotonics,
opto-mechanics and phonon transport in nanostructured devices.

\emph{We acknowledge the financial support by the European Research
Council (ERC) Starting Grant PICSEN (grant no. 306387) and the
German Research Foundation (DFG) within the Collaborative Research
Center SFB 787, the German Federal Ministry of Education and
Research (BMBF) through the VIP-project QSOURCE (Grant No. 03V0630),
and by the project EMPIR 14IND05 MIQC2 within the European Union's
Horizon 2020 research and innovation programme.}

%\bibliography{IOP,mulens}
\providecommand{\latin}[1]{#1} \makeatletter \providecommand{\doi}
  {\begingroup\let\do\@makeother\dospecials
  \catcode`\{=1 \catcode`\}=2\doi@aux}
\providecommand{\doi@aux}[1]{\endgroup\texttt{#1}} \makeatother
\providecommand*\mcitethebibliography{\thebibliography} \csname
@ifundefined\endcsname{endmcitethebibliography}
  {\let\endmcitethebibliography\endthebibliography}{}

\newpage

\begin{center}
{\bf \large Supporting Information}
\end{center}

\setcounter{figure}{0}

\renewcommand{\figurename}{Supplementary Figure}
\renewcommand{\thefigure}{S\arabic{figure}}

\section{Spectrometer response function}

The measured spectrometer response function $S(t)$ is shown in
Supplementary Figure\,\ref{fig:FigS0} together with the fit function
\begin{equation}
S(t)\! =\! 8288\cdot \exp\!\left(\!-\frac{|t-\tau_{\rm R}|}{51\,{\rm
ps}}\right) \!+\! 7888\cdot \exp\!\left[\!-\left(\frac{t-\tau_{\rm
R}}{76\,{\rm ps}}\right)^2\right]\, .
\end{equation}

\begin{figure}[h]
\centering
\includegraphics[width=0.9\columnwidth]{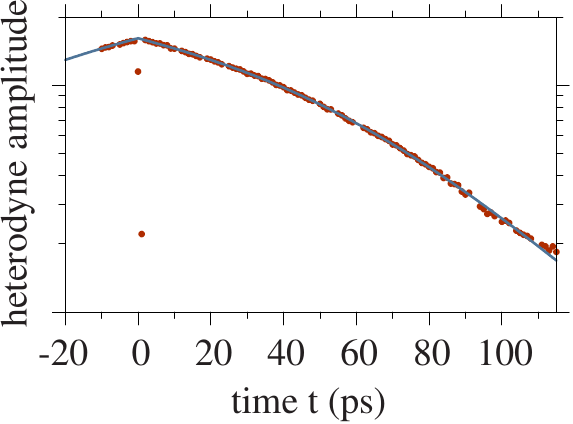}
\caption{Measured spectrometer response function with fit.
\label{fig:FigS0}}
\end{figure}

\section{Theoretical fit of the initial decay in FWM and phonon assisted transitions in PL}

The theoretical curves have been calculated on the basis of a
two-level system using a generating function formalism. The FWM
signal is obtained from the third-order nonlinear polarization given
by Eqs.~(5) and (6) in Ref.~\cite{VagovPRB03}. The PL spectra are
calculated from the linear polarization given by Eqs.~(12) and (13)
in Ref.~\cite{VagovPRB03}. The imaginary part of the Fourier
transform of this polarization gives the absorption spectrum; the PL
spectrum is then obtained as the mirror image with respect to the
ZPL.

The wave functions $\psi^{\rm i}$ for electron (i$=$e) and hole
(i$=$h) in a lens-shaped QD with harmonic confinement potential read
\begin{equation}
\psi^{\rm i} = \frac{1}{\pi^{3/4} a^{\rm i}_{r}\sqrt{a^{\rm i}_{z}}}
                \exp \left\{-\frac12\left[\left(\frac{r}{a^{\rm i}_{r}}\right)^2+\left(\frac{z}{a^{\rm i}_{z}}\right)^2\right]\right\}\ ,%\quad \text{with}\quad {\rm i=e,\ h}\ ,
\end{equation}
where $a_r^{\rm i}$ are the localization lengths in the in-plane
direction with $r^2=x^2+y^2$ and $a_z^{\rm i}$ the localization
lengths in the out-of-plane direction. Assuming the same confinement
potential for electrons and hole, the ratio between electron and
hole localization lengths is determined by the effective masses via
$a^{\rm h}/a^{\rm e}=(m^{\rm e}/m^{\rm h})^{1/4}=0.87$. The wave
functions enter in the matrix element for deformation potential
coupling in the following way:
\begin{eqnarray}
g_{{\bf q}} &=& \frac{q}{\sqrt{2\rho\hbar V\omega_{{\bf q}}}}
        \left( D^{\rm e}  \exp\left\{-\frac14 \Big[ \left(q_r^{} a^{\rm e}_r\right)^2 + \left(q_z^{} a^{\rm e}_z\right)^2 \Big]\right\} \right.\\
        &&\left.    \hspace{1.6cm}-  D^{\rm h}  \exp\left\{-\frac14 \Big[ \left(q_r^{} a^{\rm h}_r\right)^2 + \left(q_z^{} a^{\rm h}_z\right)^2\Big]\right\}\right). \notag
\end{eqnarray}
Here, ${\bf q}$ is the phonon wave vector, $\rho=5.37$\,g/cm$^3$ the
density, $V$ the volume and $\omega_{{\bf q}}=c_{\rm LA} |{\bf q}|$
is the LA phonon dispersion relation, with the sound velocity
$c_{\rm LA}=5115$\,m/s. The deformation potential coupling constants
are $D^{\rm e}$ and $D^{\rm h}$ for electron and holes,
respectively. As discussed in the paper, to achieve agreement
between theory and experiment over the whole temperature range we
have taken $D^{\rm e}=10.50$\,eV and $D^{\rm h}=-5.25$\,eV for the
curves shown in Figure\,\ref{fig:Fig4} of the paper. In principle,
one could also change the values of the density $\rho$ or the sound
velocity $c_{\rm LA}$ to increase the coupling constant $g_{\bf q}$.
However, such scaling involves a factor of $1/1.5^2$, which brings
these parameter well beyond the above-mentioned values, which are
well established in the literature. The results for additional
temperatures, namely $T=9$\,K, $15$\,K, $25$\,K and $40$\,K are
shown in Supplementary Figure\,\ref{fig:FigS1}. Also for these
curves we find an agreement between theory and experiment.

\begin{figure}[t]
\includegraphics[width=\columnwidth]{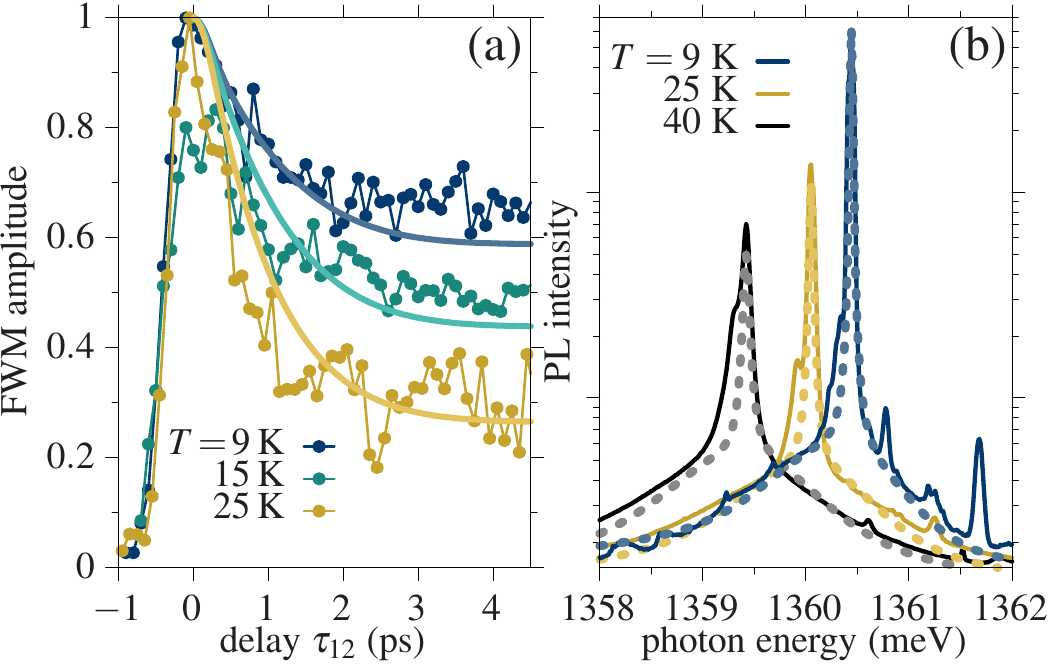}
\caption{(a) Two-pulse FWM amplitudes for small delays $\tau_{12}$;
solid lines: theoretical results. (b) PL spectra for different
temperatures; solid lines: experimental data; dashed lines:
theoretical curves (same as Figure\,\ref{fig:Fig4}b and
Figure\,\ref{fig:Fig4}c, but for temperatures $T=9$, 15, 25 and
40\,K). \label{fig:FigS1}}
\end{figure}

\begin{figure}[t]
\includegraphics[width=\columnwidth]{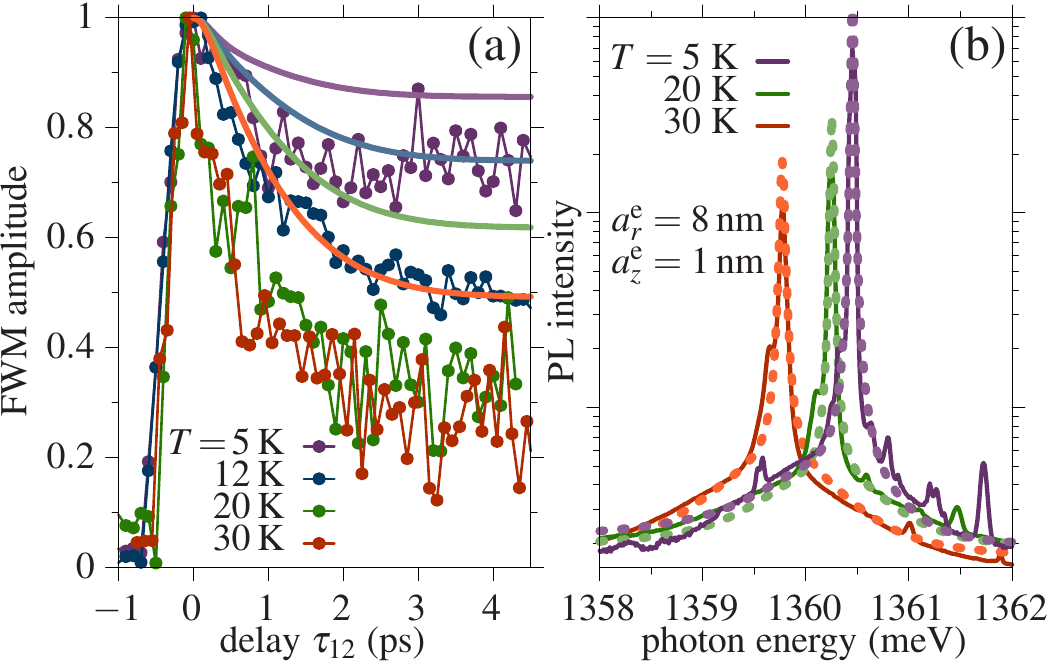}
\caption{Same as Figure\,\ref{fig:Fig4}b and
Figure\,\ref{fig:Fig4}c, but with the standard values of the
deformation potential $D^{\rm e}~\!\!=~\!\!7.0$\,eV and $D^{\rm
h}=-3.5$\,eV. \label{fig:FigS2} }
\end{figure}

Next, we discuss the theoretical fits for the standard parameters
used in several combined experimental and theoretical studies
\cite{ReiterJPC14,VagovPRB04,QuilterPRL15} with $D^{\rm e}=7.0$\,eV
and $D^{\rm h}=-3.5$\,eV when fitting both the initial FWM drop and
the PL spectra simultaneously. In this case, the only free
parameters are the localization lengths of the wave functions. In
Supplementary Figure\,\ref{fig:FigS2} we have taken $a_r^{\rm e} =
8$\,nm and $a_z^{\rm e} = 1$\,nm. For the phonon background in the
PL spectra we still get the same excellent agreement between theory
and experiment because the shape of the spectra is not influenced by
the total strength of the coupling. The height of the ZPL however is
over estimated by the simulation. This is confirmed, when looking at
the FWM signal in Supplementary Figure\,\ref{fig:FigS2}a, where we
find that the final values of the drop FWM$_{\rm drop}$, which are
reached at $\tau_{12}=5$\,ps, are different from the experimentally
observed results indicating that the coupling is obviously too weak.

\begin{figure}[t]
\includegraphics[width=\columnwidth]{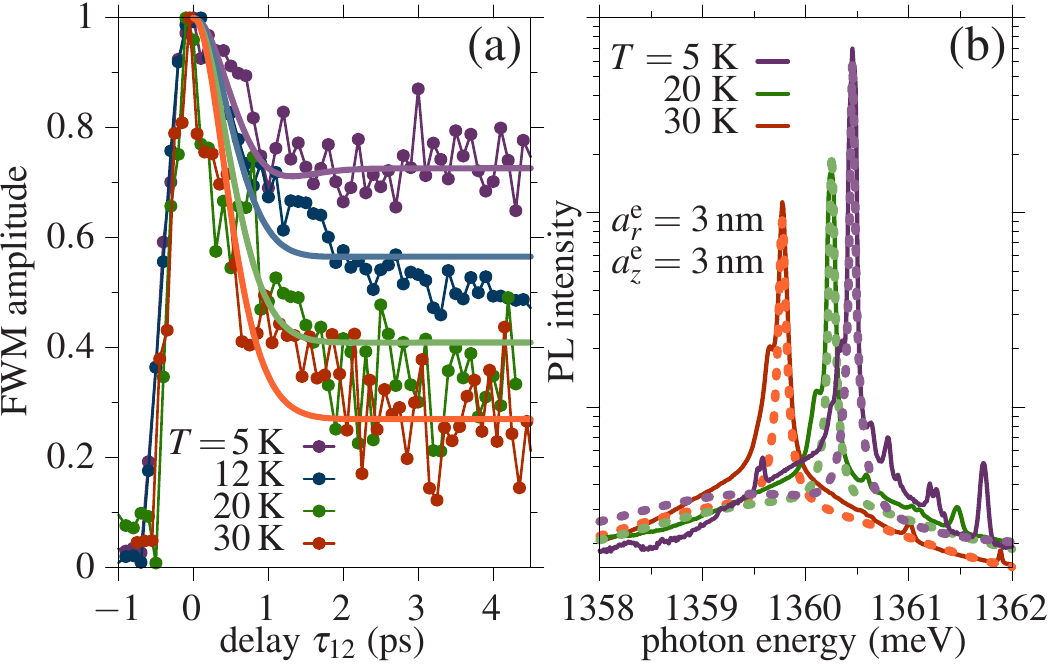}
\caption{Same as Figure\,\ref{fig:Fig4}b and
Figure\,\ref{fig:Fig4}c, but with the standard values of the
deformation potential $D^{\rm e}~\!\!=~\!\!7.0$\,eV and $D^{\rm
h}=-3.5$\,eV and with $a_r^{\rm e} = a_z^{\rm e} = 3$\,nm.
\label{fig:FigS3}}
\end{figure}

On the other hand, we can fit the initial decay in the coherence
using, e.g., a spherical dot with  $a_r^{\rm e} = a_z^{\rm e} =
3$\,nm as shown in Supplementary Figure\,\ref{fig:FigS3}. Due to the
smaller dot size the coupling is increased and we see that the final
values of the initial drop FWM$_{\rm drop}$ are now well reproduced.
However, the decay is much faster than in the case of the lens
shaped dot in Supplementary Figures\,\ref{fig:FigS1} and
\ref{fig:FigS2}. This fast drop is associated with an increased
broadening of the phonon background spectra. Accordingly, the fits
of the phonon background shown in Supplementary
Figure\,\ref{fig:FigS3}b exhibit no agreement with the experimental
measurements, while the height of the ZPL is well reproduced. Also
for other combinations of in-plane and out-of-plane localization
lengths no simultaneous fit of the spectral and temporal data could
be found.

%\bibliography{IOP,mulens}

\providecommand{\latin}[1]{#1} \makeatletter \providecommand{\doi}
  {\begingroup\let\do\@makeother\dospecials
  \catcode`\{=1 \catcode`\}=2\doi@aux}
\providecommand{\doi@aux}[1]{\endgroup\texttt{#1}} \makeatother
\providecommand*\mcitethebibliography{\thebibliography} \csname
@ifundefined\endcsname{endmcitethebibliography}
  {\let\endmcitethebibliography\endthebibliography}{}

\end{document}